\documentclass{article}
\usepackage{spconf,amsmath,graphicx}

\graphicspath{{figs/}}
\DeclareGraphicsExtensions{.png}
\usepackage{multirow}
\usepackage{arydshln}

\usepackage{bm}
\usepackage{amssymb}
\usepackage{times}
\usepackage[varg]{txfonts}
\usepackage{mathptmx}

\newcommand{\win}{{\text{W}}}
\newcommand{\shift}{{\text{S}}}
\newcommand{\margin}{{\text{M}}}

\title{Low-latency speaker-independent continuous speech separation}
\name{Takuya Yoshioka$^{\dag}$, Zhuo Chen$^{\dag}$\thanks{$^{\dag}$ Equally major contributions.}, Changliang Liu, Xiong Xiao, Hakan Erdogan, Dimitrios Dimitriadis}
\address{Microsoft, One Microsoft Way, Redmond, WA, USA}
\begin{document}
\ninept

\maketitle
\begin{abstract}
Speaker independent continuous speech separation (SI-CSS) is a task of converting a continuous audio stream, 
which may contain overlapping voices of unknown speakers, into 
a fixed number of continuous signals each of which contains no overlapping speech segment. 
A separated, or cleaned, version of each utterance is 
generated from one of SI-CSS's output channels nondeterministically without being split up and distributed to multiple channels. 
A typical application scenario is transcribing multi-party conversations, such as meetings, recorded with microphone arrays. The output signals can be simply sent to a speech recognition engine because they do not include speech overlaps. 
The previous SI-CSS method uses a neural network trained with permutation invariant training and a data-driven beamformer and thus requires 
much processing latency.
This paper proposes a low-latency SI-CSS method whose performance 
is comparable to that of the previous method 
in a microphone array-based meeting transcription task.
This is achieved (1) by using a new speech separation network architecture combined with a double buffering scheme and (2) by performing enhancement with a set of fixed beamformers followed by a neural post-filter.

\end{abstract}
\section{Introduction}
\label{sec: intro}

Overlapping speech is omnipresent in natural human-to-human conversations. 
Yet it presents a significant challenge to the current speech recognition systems, which assume an input acoustic signal to consist of up to one speaker's voice at 
every time instance. 
This work investigates the problem of recognizing human-to-human conversations which may include overlapping voices 
by using a meeting transcription task. 
We assume a microphone array to be used for audio capturing. 
The number of conversation participants is not known in advance.

Speech separation, whose goal is to untangle a mixture of co-occuring speech signals, could potentially solve the overlapping speech problem in far-field conversation transcription.
A variety of speech separation methods have been proposed in the past quater-century, ranging from 
independent component or vector analysis~\cite{Makino07}, nonnegative matrix or tensor factorization~\cite{Ozerov10}, time-frequency (TF) bin clustering~\cite{Sawada11} to deep learning~\cite{Hershey16,Drude17}. 
While considerable progress has been made, a far-field conversation transcription system that can handle speech overlaps has yet to be realized. Almost all existing speech separation methods operate on pre-segmented utterances. This requires yet another problem to be solved: speech segmentation, the goal of which is to trim each utterance from an input audio stream even when the utterance is overlapped by other voices. 
Many separation methods further assume the number of active speakers to be known beforehand, which does not hold in practice.

Speaker-independent continuous speech separation (SI-CSS)\footnote{\cite{Yoshioka18b} referred to CSS as unmixing transduction. In this paper, we use the term CSS as we feel it is more intuitive.} was proposed in \cite{Yoshioka18b} to avoid these problems. 
The idea is that, given a continuous audio stream, we want to generate
a fixed number of time-synchronous separated signals as illustrated
in Fig.~1. 
Each utterance constituting the input audio ``spurts'' from one of the output channels. When the number of active speakers is 
fewer than that of the output channels, the extra channels generate zero-valued signals.
Thus, by performing speech recognition for each separated signal, a word transcription of the entire input conversation is obtained whether it contains speech overlaps or not. 
This approach was shown to work well for meeting audio~\cite{Yoshioka18b}, outperforming the state-of-the-art data-driven beamformer using neural mask estimation~\cite{Heymann15,Boeddeker18}. 

\begin{figure}
\centering
\includegraphics[scale=0.4]{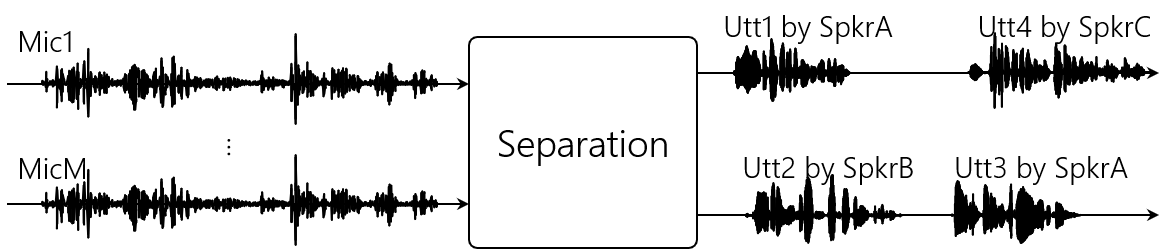}
\caption{Speaker-independent continuous speech separation.}
\label{fig: ut}
\vspace{-1em}
\end{figure}

This paper proposes a new SI-CSS method that runs with lower latency than the previous method.
Two new components are introduced to achieve the low latency processing. 
Firstly, instead of a previous bidirectional model, we employ a new separation network architecture that has recurrent connections in the forward direction and performs fixed-length look-ahead using dilated convolution.
Secondly, the segment-based data-driven beamformer of the previous method is replaced by a set of fixed beamformers followed by neural post-filtering. The post-filter removes interfering voices that remain in the beamformed signal. This is necessary as the fixed beamformers cannot precisely filter out interfering point-source signals, or other speakers' voices. 
The new method is shown to work comparably to the method of \cite{Yoshioka18b} in a meeting transcription task while requiring much lower processing latency.
A novel sound source localization (SSL) method based on a complex angular central Gaussian (cACG) distribution~\cite{Ito16} is also described.

\begin{figure*}
\centering
\includegraphics[scale=0.45]{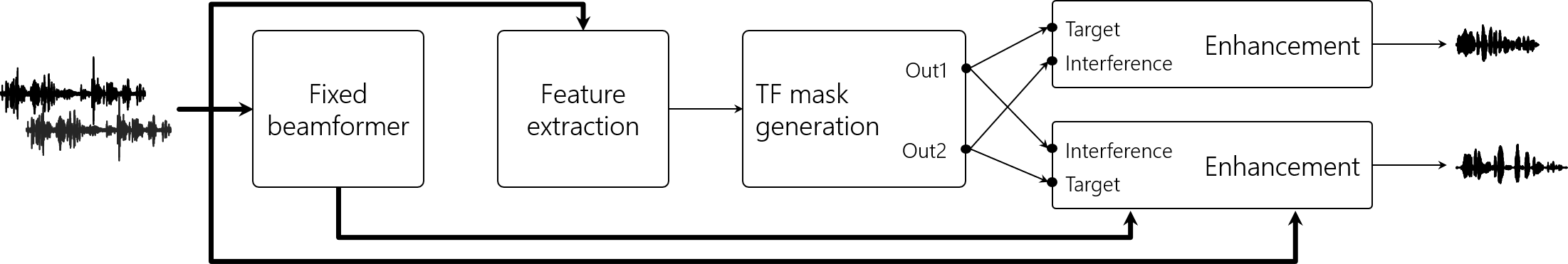}
\caption{Processing flow diagram of proposed method. Upper enhancement module also receives microphone array and beamformed signals as input. Thick lines represent multi-channel signals.}
\label{fig: overall}
\vspace{-1em}
\end{figure*}

\section{Speaker-independent\\continuous speech separation}
\label{sec: task}

This section defines the SI-CSS task and briefly reviews the method proposed in \cite{Yoshioka18b}.
The goal of SI-CSS is to transform an input signal, which may last for hours, 
into a fixed number of signals so that 
each output signal does not have overlapping speech segments. 
In this paper, we set the number of the output channels to two because three or more people rarely speak simultaneouly in meetings except for laughter segments~\cite{Cetin06}. 
A rigorous definition of the task can be found in \cite{Yoshioka18b}.
SI-CSS greatly facilitates transcribing conversations that include speech overlaps 
because we only have to perform speech recognition for each separated signal. 

The method proposed in \cite{Yoshioka18b} achieves SI-CSS as follows. 
Firstly, single- and multi-channel features are extracted from an input microphone array signal. 
The features include magnitude spectra of an arbitrarily chosen reference microphone and 
inter-microphone phase differences (IPDs)~\cite{Yoshioka18,Wang18a}. 
The stream of the feature vectors are chopped up into short segments by using a $T_\win$-second sliding window with a constant  shift of $T_\shift$ seconds.
For each segment, the extracted feature vectors are passed to a speech separation neural network that generates three TF masks: two for speech, one for noise. 
Such a network can be trained with permutation invariant training (PIT)~\cite{Kolbaek17}. 
The generated TF masks are used to construct two MVDR beamformers, each yielding a distinct separated signal. 
The beamformers are constructed by using the TF masks in a data dependent way~\cite{Heymann15,Yoshioka15b}.
In order for the beamformers to make use of a certain amount of future acoustic context so that 
the separation performance does not degrade at the end of each segment, 
the last $T_\margin$ second-part of each segment is discarded. 
Finally, the order of the separated signals are flipped if necessary to keep the output signal order consistent across segments. 

The processing latency of the method of \cite{Yoshioka18b} is $T_\shift + T_\margin + \alpha T_\win$ seconds, where $\alpha$ is the real time factor (RTF) to process
each $T_{\win}$-second segment. 
While future hardware and algorithmic improvements may reduce the RTF factor, $\alpha (<1)$, to some extent, the fixed cost of $T_\shift + T_\margin$ shall inevitably remain. 
Our latest experimental configuration sets 
$T_\shift$, $T_\margin$, and $T_\win$ at 0.8, 0.4, and 2.4, respectively, which reasonably balances the separation performance and the computational cost.


\section{Proposed Method}
\label{sec: proposed}

Figure \ref{fig: overall} illustrates the processing flow of the proposed method. 
Firstly, magnitude spectra and IPD features are extracted from an input multi-channel signal. 
They are fed to a TF mask generation module, which is implemented by using a neural network trained with a mean squared error (MSE) PIT loss as with 
the previous method (see \cite{Yoshioka18b} for details). 
The TF mask generation module \textit{continuously} yields two sets of TF masks with a small time lag. 
While the TF masks may be applied directly to the input signal, direct masking tends to end up with degrading speech recognition performance due to speech distortion. 
Thus, we fed them to another system component, referred to as an enhancement module in Fig.~\ref{fig: overall}, which utilizes fixed beamformers and a neural network-based post-filter. 
The rest of this section details each component other than the enhancement module, which we elaborate on in the next section.

\subsection{Fixed beamformers}

For real time applications, beamformers designed for a specific microphone array geometry are more advantageous than the data-driven beamforming approach~\cite{Heymann15,Yoshioka15b}.  It is noteworthy that, as demonstrated in \cite{Boeddeker18}, a well-designed fixed beamformer is as effective at reducing background noise as the state-of-the-art data-driven beamformer.

We designed a set of 18 fixed beamformers, each with a distinct focus direction, for the seven-channel circular microphone array that we used for our data collection. 
The focus directions of neighboring beamformers are separated by 20 degrees. 
The beam pattern for each direction was optimized to maximize the output signal-to-noise ratio for simulated environments. 

\subsection{Feature extraction}

Multiple independent reports~\cite{Yoshioka18,Wang18a,Wang18c} show IPD feature's effectiveness for neural speech separation. 
In this work, we make use of both the IPDs and the magnitude spectrum of the signal of the first, or reference, microphone. 
The IPDs are computed between the reference microphone and each of the other microphones.

\subsection{Time-frequency mask generation}

A neural network trained with PIT generates TF masks for speech separation from the features computed as described above.
The most prominent advantage of PIT over other speech separation mask estimation schemes, such as spatial clustering~\cite{Drude14,Ito14}, 
deep clustering~\cite{Hershey16}, and deep attractor networks~\cite{Chen17}, is that 
it does not require prior knowledge of the number of active speakers. 
When only one speaker is active, the PIT-trained network yields zero-valued masks from extra output channels. 
This is desirable for SI-CSS because we always generate a fixed number of output signals.

\subsubsection{Network architecture}

Prior work on PIT often utilized bidirectional models. 
A neural network trained with PIT can not only separate speech signals for each short time frame but also keep the order of the output signals consistent across short time frames. 
This is possible largely because the network is penalized if it changes the output signal order at some middle point of an utterance during training. 
On the other hand, 
for the network to be able to consistently assign an output channel to each separated signal frame, 
it is also beneficial for the network to take account of some future acoustic context~\cite{Kolbaek17}.
Therefore, bidirectional models are inherently advantageous while their use hinders low latency processing. 

\begin{figure}
\centering
\includegraphics[scale=0.45]{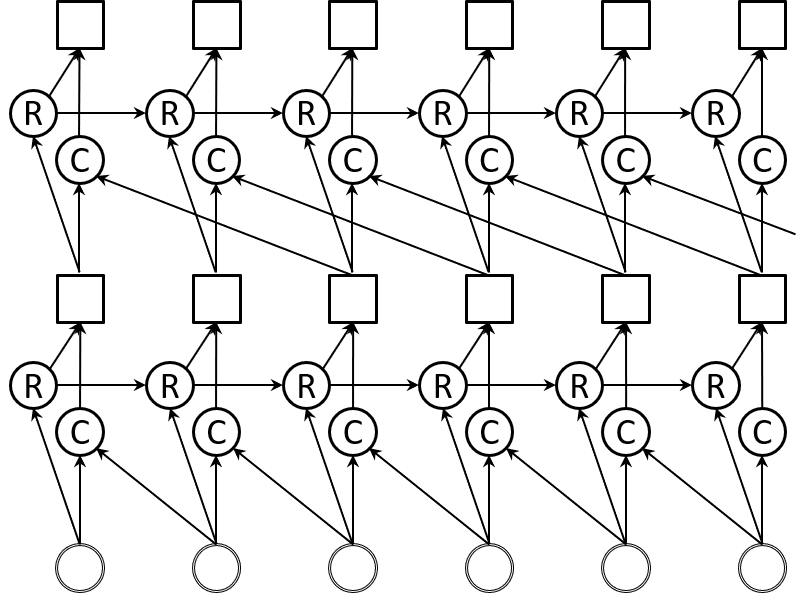}
\caption{RNN-CNN hybrid model. ``R'' and ``C'' circles represent recurrent (LSTM) and convolution nodes, respectively. Square nodes perform splicing. Double circles represent input nodes.}
\label{fig: network}
\vspace{-1em}
\end{figure}

In this paper, we propose to use a hybrid of a unidirectional recurrent neural network (RNN) and a convolutional neural network (CNN). 
Figure \ref{fig: network} depicts the architecture of our RNN-CNN hybrid model. 
The temporal acoustic dependency in the forward direction is modeled by the RNN, or more specifically a long short term memory (LSTM) network. 
On the other hand, the CNN captures the backward acoustic dependency. 
Dilated convolution~\cite{Oord16} is used as shown in Fig.~\ref{fig: network}
to efficiently cover a fixed length of future acoustic context.
Our experimental system consists of a projection layer with 1024 units, two RNN-CNN hybrid layers, and two parallel fully connected layers with sigmoid nonlinearity. 
The final layer's activations are used as TF masks for speech separation. 
With the two RNN-CNN hybrid layers, 
our model utilizes four ($=N_{\text{LF}}$) future frames, where our frame shift is 0.016 seconds. 

\subsubsection{Double buffering}

While the PIT-trained network is designed to assign an output channel to each separated speech frame consistently across short time frames, we cannot simply keep feeding
the network with the feature vectors for a long time. 
Firstly, the speech separation network is trained on mixed speech segments of up to $T_{\text{TR}} (=10)$ seconds during the learning phase. 
The resultant model does not necessarily keep the output order consistent beyond $T_{\text{TR}}$ seconds. 
In addition, RNN's state values tend to saturate after a while when it is exposed to a long feature vector stream~\cite{Chang18}. 
Therefore, the state values need to be refreshed at some interval in such a way that 
keeps the output order consistent. 

\begin{figure}
\centering
\includegraphics[scale=0.4]{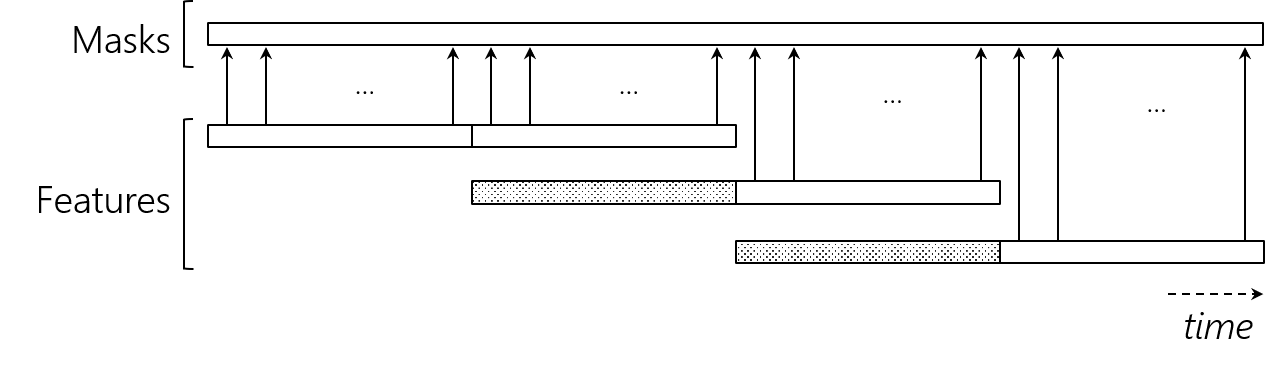}
\caption{Double buffering for real-time CSS. TF masks calculated for shaded blocks are used only for ordering output channels.}
\label{fig: buffer}
\vspace{-1em}
\end{figure}

To address this problem, we propose a double buffering scheme as illustrated in Fig~\ref{fig: buffer}.
We feed feature vectors to the network for $T_\win (=2.4)$ seconds. 
Because the model uses a fixed length of future context, the output TF masks can be obtained with a limited processing latency. 
Halfway through processing the first buffer, we start a new buffer from fresh RNN state values.  
The new buffer is processed for another $T_\win$ seconds. 
By using the TF masks generated for the first $T_\win /2$-second half, we determine the best output order for the second buffer. 
The order is determined so that the MSE can be minimized between the separated signals obtained for the last half of the previous buffer and those for the first half of the current buffer. 
By using two buffers in this way, the TF masks can be continuously generated for a long stream of audio in real time.

\begin{figure}
\centering
\includegraphics[scale=0.45]{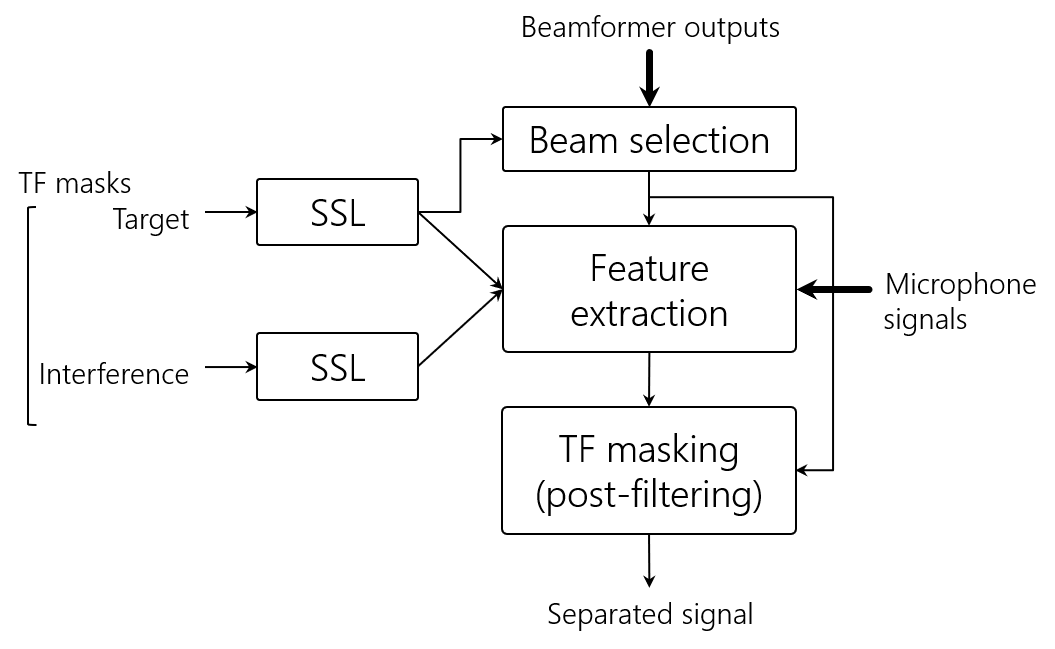}
\caption{Processing flow diagram of target speech enhancement.}
\label{fig: enhancement}
\vspace{-1em}
\end{figure}

\section{Target speech enhancement}
\label{sec: enh}

Given two TF masks, one for a target speaker and one for an interfering speaker, and multiple beamformed signals, 
the enhancement module generates a signal where the target speaker is enhanced against the interfering speaker and background noise. 
As shown in Fig.~\ref{fig: enhancement}, this is performed by first selecting the beamformer channel pointing at the target speaker direction and then post-filtering the signal
with TF masks derived from a post-filtering neural network.  
Unlike the separation network, the post-filtering network receives the target and interference angles as input in addition to the microphone and beamformed signals
in order to enhance only the target speaker's voice. Our network model does not use any future data frames.

\subsection{Sound source localization}
The enhancement processing starts with performing SSL for each of the target and interference speakers. 
The estimated directions are used both for selecting the beamformer channel and as an input to the post-filtering network. 

For computational efficiency, the target and interference directions are estimated every $N_\shift$ frames, or $0.016N_\shift$ seconds. 
For each of the target and interference, SSL is performed by using the input multi-channel audio and the TF masks in frames $(n-N_\win, n]$, where $n$ refers to the current frame index. 
The estimated directions are used for processing the frames in $(n-N_\margin - N_\shift, n-N_\margin]$, resulting in delay of $N_\margin$ frames. 
The ``margin'' of length $N_\margin$ is introduced so that SSL leverages
a small amount of future context. 
In our experiments, $N_\margin$, $N_\shift$, and $N_\win$ are set at 
20, 10, and 50, respectively. 

SSL is achieved with maximum likelihood estimation using the TF masks as observation weights.
We hypothesize that each magnitude-normalized multi-channel observation vector, $\bm{z}_{t,f}$, follows a cACG distribution~\cite{Ito16} as follows:
\begin{align}
p( \bm{z}_{t,f} | \omega ) =  0.5 \pi^{-M} (M-1)!   |\bm{B}_{f, \omega} |^{-1} \bigl( \bm{z}_{t,f}^H \bm{B}_{f, \omega}^{-1} \bm{z}_{t,f} \bigr)^{-M}, 
\end{align}
where $\omega$ denotes an incident angle, $M$ the number of microphones, and $\bm{B}_{f, \omega} = ( \bm{h}_{f,\omega} \bm{h}_{f,\omega}^H + \epsilon \bm{I})$ with $\bm{h}_{f, \omega}$, $\bm{I}$, and $\epsilon$ being the steering vector for angle $\omega$, the $M$-dimensional identify matrix, and a small flooring value.
Given a set of observations, $  Z = \{ \bm{z}_{t,f}  \} $, we want to maximize the following log likelihood function with respect to $\omega$: 
\begin{align}
L(\omega) = \sum_{t,f} m_{t,f} \log p( \bm{z}_{t,f} |  \omega ), 
\end{align}
where $\omega$ can take a discrete value in $[0, 360)$ and $m_{t,f}$ denotes the TF mask provided by the separation network. 
It can be shown that the log likelihood function reduces to the following simple form: 
\begin{align}
L(\omega) = - \sum_{t,f} m_{t,f} \log \bigl(  1 - ||  \bm{z}_{t,f}^H\bm{h}_{f,\omega} ||^2 / (1+\epsilon) \bigr).
\end{align}
$L(\omega)$ is computed for every possible discrete angle value. The $\omega$ value that gives the highest score is picked as a direction estimate.
Further analysis of the cACG-based SSL method will be conducted 
in a separate paper.

\subsection{Neural post-filtering}

The beamformer signal selected based on the estimated target speaker's direction is further processed with TF masking. The aim is to cancel 
the interfering speaker's voice that has been left to the beamformed signal. 
This post-filtering is indispensable because fixed beamformers are usually designed to remove diffuse noise and thus cannot remove interfering speech signals effectively. 

For this purpose, we employ the direction-informed target speech extraction method proposed in \cite{Chen18}. The method uses a neural network
that accepts features computed based on the target and interference directions 
to focus on the target direction and give less attention to the interference direction. 
The network generates TF masks that can extract only the target speaker component from the 
input beamformed audio. 
The directional feature is calculated for each TF bin as a sparsified version of the cosine distance between the target direction's steering vector and the microphone array signal. The IPD features and the magnitude spectrum of the beamformed signal are also fed to the network. 
The model consists of four uni-directional LSTM layers, each with 600 units, and is trained to minimize the MSE of clean and TF mask-processed signals. 
We refer the reader to \cite{Chen18} for further details. 

In summary, the minimum processing latency required for executing the proposed method is 
$N_{\text{LF}} + N_\margin$ frames, where the frame shift is 0.016 seconds.
In our experiments, the look-ahead size, $N_{\text{LF}}$, of the RNN-CNN hybrid model is four while $N_\margin$ is set at 20. This is much smaller than the lower-bound latency of the previous method, i.e., $T_\shift + T_\margin$ seconds.

\section{Experiments}
\label{sec: exp}

We conducted meeting speech recognition experiments to evaluate the effectiveness of the proposed SI-CSS method. We performed SI-CSS on multi-microphone meeting recordings and sent the separated signals to a speech recognition engine to obtain 
word transcriptions. 
The results were scored with asclite tool~\cite{Fiscus06}, which aligns multiple (two for our work) hypotheses against multiple speaker-specific reference transcriptions to
generate word error rate (WER) estimates. 

We recorded and transcribed six meetings at our Speech Group. Both headset microphones and a seven-channel circular microphone array were used. 
Our meetings were conducted at multiple conference rooms. 
The number of the meeting attendees varied from four to eleven as shown in Table~\ref{tab: result}.

Our separation network was trained on 600 hours of artificially reverberated and mixed speech signals while 
the post-filter network was trained on 1.5K hours of data. 
See \cite{Yoshioka18b,Chen18} for our simulation and training procedures. 
Multi-channel dereverberation is performed prior to SI-CSS in real time by using the weighted prediction error (WPE) method~\cite{Yoshioka12c}. 
Our acoustic model was sequence-trained on 33K hours of audio, including artificially contaminated speech.
Decoding was performed with a trigram language model.

\subsection{Results}

\begin{table}
\centering
\vspace{.5em}
\caption{\%WER of different methods for meeting transcription. Numbers of meeting attendees shown in parentheses. FBF: fixed beamformer; PF: post-filter.}
\label{tab: result}
\begin{tabular}{|l||c|c|c|}\hline
System     & S1 & S2 & S3 (Proposed) \\ \hline\hline
Sep. model & BLSTM & R/CNN hybrid & R/CNN hybrid \\ \hline
Enh. method & MVDR & MVDR & FBF-PF \\ \hline\hline
MTG0 (4) & 22.1 & 23.7 & 20.7 \\ \hline
MTG1 (6) & 17.0 & 18.1 & 18.0 \\ \hline
MTG2 (6) & 20.6 & 21.8 & 22.0 \\ \hline
MTG3 (8) & 28.0 & 27.8 & 28.4 \\ \hline
MTG4 (4) & 28.5 & 29.8 & 29.5 \\ \hline
MTG5 (11) & 21.0 & 22.9 & 20.5 \\ \hline\hline
Overall & 21.5 & 22.7 & 21.7 \\ \hline
\end{tabular}
\vspace{-1em}
\end{table}

Table \ref{tab: result} lists the WERs of the previous method (S1) and the proposed method (S3). The performance of a system that yields separated signals by using MVDR and the RNN-CNN hybrid model is also presented (S2). The performance of the proposed method is comparable to that of the previous method. Comparison of S1 and S2 reveals that the use of the RNN-CNN hybrid model slightly degraded the quality of the speech separation masks. The proposed enhancement scheme, combining the fixed beamformers with the post-filter, was less sensitive to the degradation in the TF mask quality. 
This would be because the separation TF masks are used only for SSL in the proposed method while data-driven MVDR significantly relies on the TF masks. 

\begin{table}
\centering
\caption{Impact of SSL window configurations.}
\label{tab: window_effect}
\begin{tabular}{|l||c|c|c|}\hline
Window size ($N_\win$) & 50 & 50 & 70 \\ \hline
Margin ($N_\margin$) & 20 & 10 & 30 \\ \hline\hline
\%WER & 21.7 & 22.4 & 21.7 \\ \hline
\end{tabular}
\vspace{-1em}
\end{table}

Table \ref{tab: window_effect} compares the WERs for different SSL window configurations. It can be seen that having a certain number of margin frames has non-negligible impact on the separation performance. 
A margin of 20 frames, or 0.32 seconds, seems sufficient to achieve the performance on par with the previous method using a bidirectional model and data-driven MVDR beamforming.

\section{Conclusion}
\label{sec: conclusion}

In this paper, we described a novel low-latency SI-CSS method which 
uses an RNN-CNN hybrid network for generating speech separation TF masks
and a set of fixed beamformers followed by a neural post-filter. 
A double buffering scheme is introduced to continuously generate the TF masks with a short amount of delay. 
A new maximum likelihood SSL method using a cACG model is also presented. 
The proposed method achieved comparable meeting transcription accuracy to that of the previously proposed method while significantly reducing the processing latency.

\vfill\pagebreak

\bibliographystyle{IEEEbib}
\bibliography{ICASSP2019}

\end{document}